\documentclass[twocolumn,aps,prl,floatfix,superscriptaddress]{revtex4-1}
\usepackage{amsmath,amssymb}
\usepackage{graphicx,float}
\usepackage{times,txfonts}
\usepackage{color}
\usepackage{multirow}
\usepackage{bbold}
\usepackage{color}
\usepackage{soul}
\usepackage{hyperref}
\usepackage{times,txfonts}
\usepackage{natbib}
\usepackage{blindtext}
\usepackage[export]{adjustbox}
\usepackage{mathrsfs}
\usepackage{textcomp}
\usepackage{blkarray}
\usepackage{multirow}

\newcommand{\bra}[1]{\left\langle #1 \right|}
\newcommand{\ket}[1]{\left| #1 \right\rangle}
\newcommand{\braket}[2]{\left\langle {#1{\left| \vphantom{#1 #2} \right.} #2} \right\rangle}

\renewcommand{\epsilon}{\varepsilon}
\renewcommand{\phi}{\varphi}

\def\VR{\kern-\arraycolsep\strut\vrule &\kern-\arraycolsep}
\def\vr{\kern-\arraycolsep & \kern-\arraycolsep}
\usepackage{sistyle}
\usepackage{soul}
\definecolor{lightblue}{RGB}{185,210,248}
\sethlcolor{lightblue}

\begin{document}
%
\title{High-Dimensional Intra-City Quantum Cryptography with Structured Photons}
\author{Alicia Sit}
\affiliation{Physics Department, Centre for Research in Photonics, University of Ottawa, Advanced Research Complex, 25 Templeton, Ottawa ON Canada, K1N 6N5}
\author{Fr\'ed\'eric Bouchard}
\affiliation{Physics Department, Centre for Research in Photonics, University of Ottawa, Advanced Research Complex, 25 Templeton, Ottawa ON Canada, K1N 6N5}
\author{Robert Fickler}
\affiliation{Physics Department, Centre for Research in Photonics, University of Ottawa, Advanced Research Complex, 25 Templeton, Ottawa ON Canada, K1N 6N5}
\author{J\'er\'emie Gagnon-Bischoff}
\affiliation{Physics Department, Centre for Research in Photonics, University of Ottawa, Advanced Research Complex, 25 Templeton, Ottawa ON Canada, K1N 6N5}
\author{Hugo Larocque}
\affiliation{Physics Department, Centre for Research in Photonics, University of Ottawa, Advanced Research Complex, 25 Templeton, Ottawa ON Canada, K1N 6N5}
\author{Khabat Heshami}
\affiliation{National Research Council of Canada, 100 Sussex Drive, Ottawa ON Canada, K1A 0R6}
\author{Dominique Elser}
\affiliation{Max-Planck-Institut f\"ur die Physik des Lichts, G\"{u}nther-Scharowsky-Stra{\ss}e 1,Bau 24, 91058 Erlangen, Germany}
\affiliation{Institut f\"ur Optik, Information und Photonik, Universit\"{a}t Erlangen-N\"{u}rnberg, Staudtstra{\ss}e 2, 91058 Erlangen, Germany}
\author{Christian Peuntinger}
\altaffiliation[Current address: ]{Department of Physics, University of Otago, 730 Cumberland Street, Dunedin 9016, New Zealand}
\affiliation{Max-Planck-Institut f\"ur die Physik des Lichts, G\"{u}nther-Scharowsky-Stra{\ss}e 1,Bau 24, 91058 Erlangen, Germany}
\affiliation{Institut f\"ur Optik, Information und Photonik, Universit\"{a}t Erlangen-N\"{u}rnberg, Staudtstra{\ss}e 2, 91058 Erlangen, Germany}
\author{Kevin G\"{u}nthner}
\affiliation{Max-Planck-Institut f\"ur die Physik des Lichts, G\"{u}nther-Scharowsky-Stra{\ss}e 1,Bau 24, 91058 Erlangen, Germany}
\affiliation{Institut f\"ur Optik, Information und Photonik, Universit\"{a}t Erlangen-N\"{u}rnberg, Staudtstra{\ss}e 2, 91058 Erlangen, Germany}
\author{Bettina Heim}
\altaffiliation[Current address: ]{OHB System AG, Manfred-Fuchs-Stra{\ss}e 1, 82234 We{\ss}ling}
\affiliation{Max-Planck-Institut f\"ur die Physik des Lichts, G\"{u}nther-Scharowsky-Stra{\ss}e 1,Bau 24, 91058 Erlangen, Germany}
\affiliation{Institut f\"ur Optik, Information und Photonik, Universit\"{a}t Erlangen-N\"{u}rnberg, Staudtstra{\ss}e 2, 91058 Erlangen, Germany}
\author{Christoph Marquardt}
\affiliation{Max-Planck-Institut f\"ur die Physik des Lichts, G\"{u}nther-Scharowsky-Stra{\ss}e 1,Bau 24, 91058 Erlangen, Germany}
\affiliation{Institut f\"ur Optik, Information und Photonik, Universit\"{a}t Erlangen-N\"{u}rnberg, Staudtstra{\ss}e 2, 91058 Erlangen, Germany}
\author{Gerd Leuchs}
\affiliation{Physics Department, Centre for Research in Photonics, University of Ottawa, Advanced Research Complex, 25 Templeton, Ottawa ON Canada, K1N 6N5}
\affiliation{Max-Planck-Institut f\"ur die Physik des Lichts, G\"{u}nther-Scharowsky-Stra{\ss}e 1,Bau 24, 91058 Erlangen, Germany}
\affiliation{Institut f\"ur Optik, Information und Photonik, Universit\"{a}t Erlangen-N\"{u}rnberg, Staudtstra{\ss}e 2, 91058 Erlangen, Germany}
\author{Robert W. Boyd}
\affiliation{Physics Department, Centre for Research in Photonics, University of Ottawa, Advanced Research Complex, 25 Templeton, Ottawa ON Canada, K1N 6N5}
\affiliation{Institute of Optics, University of Rochester, Rochester, New York, 14627, USA}
\author{Ebrahim Karimi}
\affiliation{Physics Department, Centre for Research in Photonics, University of Ottawa, Advanced Research Complex, 25 Templeton, Ottawa ON Canada, K1N 6N5}
\affiliation{Department of Physics, Institute for Advanced Studies in Basic Sciences, 45137-66731 Zanjan, Iran}
\maketitle

\noindent\textbf{Quantum key distribution (QKD) promises information-theoretically secure communication, and is already on the verge of commercialization. Thus far, different QKD protocols have been proposed theoretically and implemented experimentally~\cite{scarani2009security,lo2014secure}. The next step will be to implement high-dimensional protocols in order to improve noise resistance and increase the data rate~\cite{bechmann2000quantum,cerf2002security,groblacher2006experimental,mafu2013higher,mirhosseini2015high}. Hitherto, no experimental verification of high-dimensional QKD in the single-photon regime has been conducted outside of the laboratory. Here, we report the realization of such a single-photon QKD system in a turbulent free-space link of 0.3~km over the city of Ottawa, taking advantage of both the spin and orbital angular momentum photonic degrees of freedom. This combination of optical angular momenta allows us to create a 4-dimensional state~\cite{nagali2010experimental}; wherein, using a high-dimensional BB84 protocol~\cite{bechmann2000quantum,cerf2002security}, a quantum bit error rate of 11\% was attained with a corresponding secret key rate of 0.65 bits per sifted photon. While an error rate of 5\% with a secret key rate of 0.43 bits per sifted photon is achieved for the case of 2-dimensional structured photons. Even through moderate turbulence without active wavefront correction, it is possible to securely transmit information carried by structured photons, opening the way for intra-city high-dimensional quantum communications under realistic conditions.}

In addition to wavelength and polarization, a light wave is characterized by its orbital angular momentum (OAM)~\cite{allen1992orbital}, which corresponds to its helical wavefronts. Polarization is naturally bi-dimensional, i.e. $\{ \ket{L},\ket{R}\}$, and the associated angular momentum can take the values of $\pm\hbar$ per photon, where $\hbar$ is the reduced Planck constant, and $\ket{L}$ and $\ket{R}$ are left- and right-handed circular polarizations, respectively. In contrast, OAM is inherently unbounded, such that a photon with $\ell$ intertwined helical wavefronts, $\ket{\ell}$, carries $\ell\hbar$ units of OAM, where $\ell$ is an integer~\cite{mair2001entanglement}. Quantum states of light resulting from an arbitrary coherent superposition of different polarizations and spatial modes, e.g. OAM, are referred to as \emph{structured photons}; these photons can be used to realize higher-dimensional states of light~\cite{nagali2010experimental}. Aside from their fundamental significance in quantum physics~\cite{molina:2007,cardano2015quantum}, single photons encoded in higher dimensions provide an advantage in terms of security tolerance and encrypting alphabets for quantum cryptography~\cite{bechmann2000quantum,cerf2002security,mirhosseini2015high} and classical communications~\cite{willner2015optical}. The behaviour of light carrying OAM through turbulent conditions has been studied theoretically and simulated in the laboratory scale~\cite{paterson2005atmospheric,malik2012influence,farias2015resilience,goyal2016effect}.
Experimentally, OAM states have been tested in classical communications across intra-city links in Los Angeles (120~m)~\cite{wang2012terabit}, Venice (420~m)~\cite{tamburini:2012}, Erlangen (1.6~km)~\cite{lavery:2015}, Vienna (3~km)~\cite{krenn:2014}, and between two Canary Islands (143~km)~\cite{Krenn2016Twisted} which is the longest link thus far. With attenuated lasers, OAM states and vector vortex beams have been respectively implemented in high-dimensional and 2-dimensional BB84 protocols, where the former was performed in a laboratory~\cite{mirhosseini2015high}, and the latter in a hall in Padua (210~m)~\cite{vallone:2014}. Though not QKD, entanglement distribution of bi-dimensional twisted photons has been recently studied across the Vienna link~\cite{krenn2015twisted}.

%
\begin{figure}[h]
\begin{center}
\includegraphics[width=0.9\columnwidth]{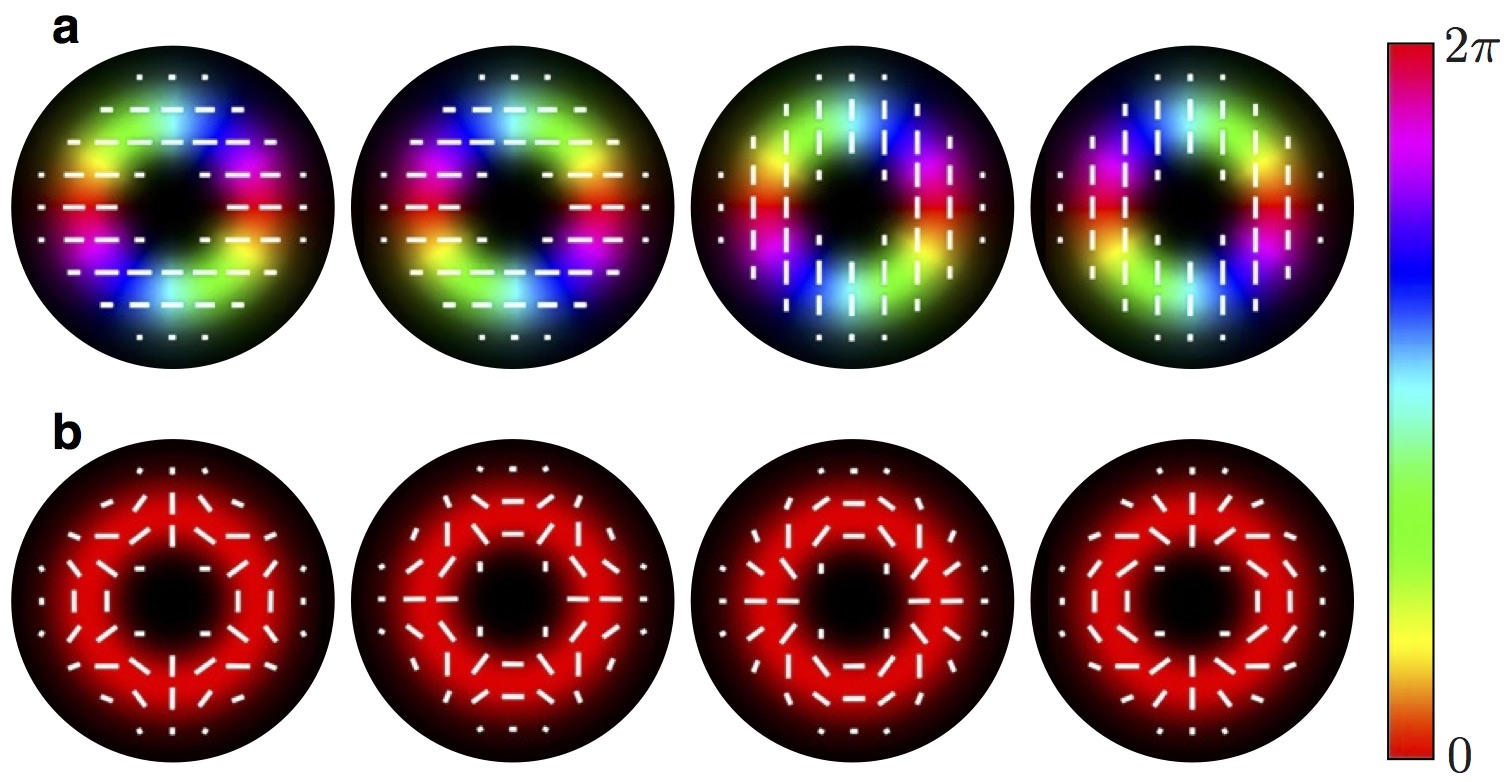}
\caption[]{{\bf Mode structure of mutually unbiased bases for $\boldsymbol{\ell=2}$}. {\bf a}, $\{\ket{\psi}^i\}$ and {\bf b}, $\{\ket{\phi}^j\}$ are examples of two bases of structured states of light, encoding in both polarization and OAM of $\ell=2$. Each basis is orthonormal, and the two bases are mutually unbiased with respect to each other such that $|^i\!\braket{\psi}{\phi}^j|^2=1/4$. These MUBs have the advantage of possessing identical intensity profiles --- ``doughnut'' shaped --- and are shape-invariant upon free-space propagation. The information, therefore, is encoded in the transverse polarization and phase distributions, denoted by the white lines and hue colour, respectively.}
\label{fig:fig1}
\end{center}
\end{figure}

%
\begin{figure*}
	\begin{center}
	\includegraphics[width=0.95\textwidth]{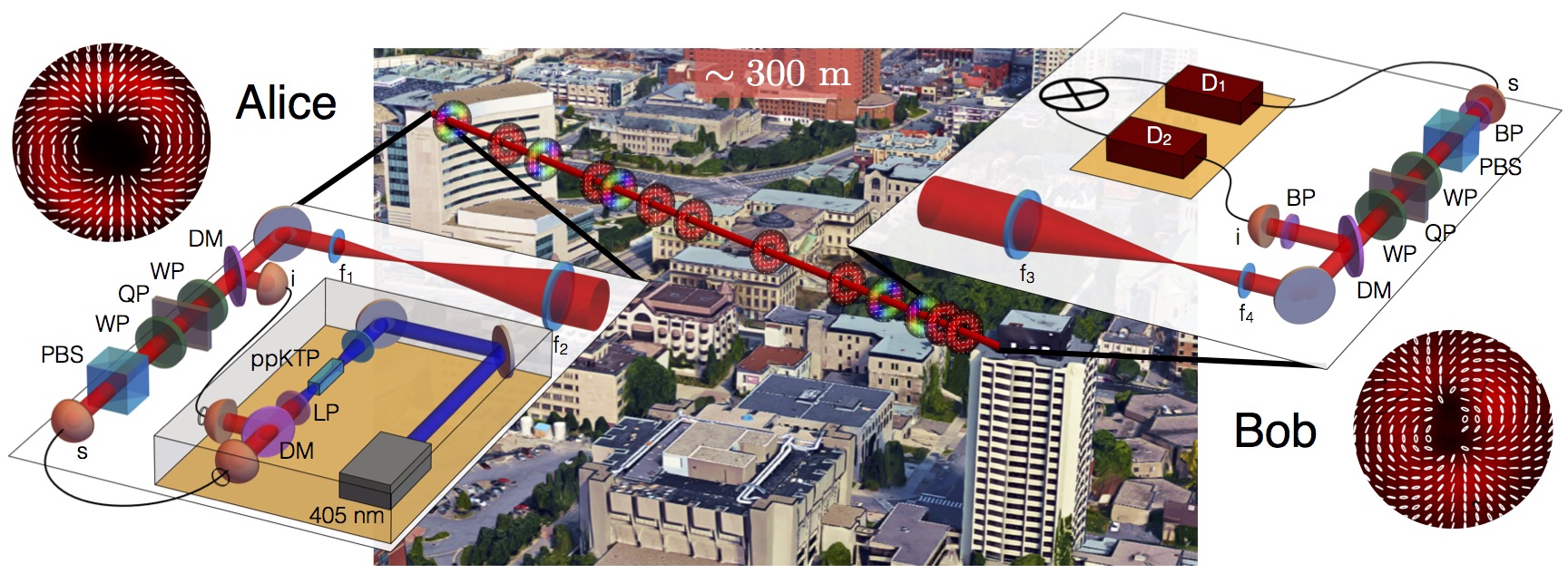}
	\caption[]{{\bf Ottawa intra-city quantum communication link.} Schematic of sender (left) with heralded single-photon source (signal, s, and idler, i) and Alice's state preparation setup. Alice prepares a state from $\{\ket{\psi}^i\}$ or $\{\ket{\phi}^j\}$ using a polarizing beam-splitter (PBS), waveplates (WP), and $q$-plate (QP). The signal and idler photons are recombined on a dichroic mirror (DM) before being sent to Bob. Two telescopes comprised of lenses with focal lengths of $f_1=75$~mm, $f_2=f_3= 400$~mm (diameter of 75~mm), and $f_4=50$~mm are used to enlarge and collect the beam, minimizing its divergence through the 0.3~km link. Bob, receiver (right), can then perform measurements on the sent states and record the coincidences between the signal and idler photons with detectors $D_1$ and $D_2$ at a coincidence logic box. Enclosures are built around the sender and receiver to shelter them from the wind and weather, as well as to shield them from background light. Examples of experimentally reconstructed polarization distributions for a structured mode from $\{\ket{\phi}^j\}$ using a CW laser that Alice prepared (top left) and Bob measured (bottom right) are shown in the insets. Figure legend: ppKTP = periodically-poled KTP crystal, LP=long-pass filter, BP=band-pass filter.  Map data: Google Maps, \textcopyright~2016.}
	\label{fig:fig2}
	\end{center}
\end{figure*}

In this Letter, we combine polarization $\{\ket{H}, \ket{V}\}$ and an OAM subspace of $\{ \ket{\ell}, \ket{-\ell} \}$ to form 4-dimensional quantum states $\ket{k}$, for $k$=1,2,3,4, belonging to the set $\{ \ket{H,\ell}, \ket{V,\ell}, \ket{H,-\ell}, \ket{V,-\ell} \}$, where $\ket{H}=(\ket{L}+\ket{R})/\sqrt{2}$ and $\ket{V}=-i(\ket{L}-\ket{R})/\sqrt{2}$ are horizontal and vertical polarization states, respectively. We can create two sets of mutually unbiased bases (MUBs) from $\ket{k}$, defined as $\ket{\psi}^i = \mathcal{M}_0^{ik} \ket{k}$ and $\ket{\phi}^j = \mathcal{M}_1^{jk} \ket{k}$, where $|^i\!\braket{\psi}{\psi}^j|^2=|^i\!\braket{\phi}{\phi}^j|^2=\delta_{ij}$, and $|^i\!\braket{\psi}{\phi}^j|^2=1/4$ (see Supplementary information for $\mathcal{M}_0$ and $\mathcal{M}_1$). Figure 1 illustrates the spatial structure of these MUBs for the case of $\ell=2$. The information encoded within these modes lies in the transverse polarization and phase distributions; however, all of these modes possess a ``doughnut'' shaped intensity distribution. The polarization distributions contain only linearly polarized states, and such beams are commonly called vector vortex beams~\cite{zhan2009cylindrical}; in the case of $\{\ket{\phi}^j\}$, the linear polarizations vary across the transverse plane. $\{\ket{\psi}^i\}$ and $\{\ket{\phi}^j\}$ are conjugate quantities, and based on quantum complementarity they cannot be measured simultaneously; this forms the backbone of security in quantum cryptography. Specifically, in the BB84 protocol~\cite{bennett1984quantum}, the bases of preparation and measurement are randomly chosen between two MUBs by a sender and receiver, traditionally called Alice and Bob, respectively. We used the two MUBs of structured modes, $\{\ket{\psi}^i\}$ and $\{\ket{\phi}^j\}$, to perform a high-dimensional BB84 protocol~\cite{bechmann2000quantum,cerf2002security}. There are different approaches used to generate and sort these structured modes of light. We utilize liquid crystal devices known as $q$-plates~\cite{marrucci2006optical}, which coherently couple optical spin angular momentum to OAM. $Q$-plates are advantageous as they are placed in-line, are efficient in comparison to diffractive elements, and can be used to create arbitrary complex modal structures~\cite{larocque2016arbitrary}. These $q$-plates used in conjunction with a carefully chosen sequence of waveplates can generate $\{\ket{\psi}^i\}$ and $\{\ket{\phi}^j\}$ (see Supplementary information for details). Furthermore, it is possible to rapidly switch between the states in $\{\ket{\psi}^i\}$ and $\{\ket{\phi}^j\}$, on the order of 1~MHz, by replacing the waveplates with Pockels cells. Since $q$-plates are coherent and linear devices, they also work in the single photon regime~\cite{nagali2009quantum}.

We built a free-space link between the rooftops of two buildings, 0.3~km apart and 40~m above the ground, on the University of Ottawa campus; see Fig.~\ref{fig:fig2}. Two enclosures were constructed to contain and protect all of the optics and equipment at the sender and receiver. The sender unit is comprised of both the heralded single-photon source and the setup where Alice can prepare states. The receiver unit contains Bob's state measurement setup and the single photon detection system (see Methods for experimental details).
\begin{figure*}
	\begin{center}
	\includegraphics[width=2\columnwidth]{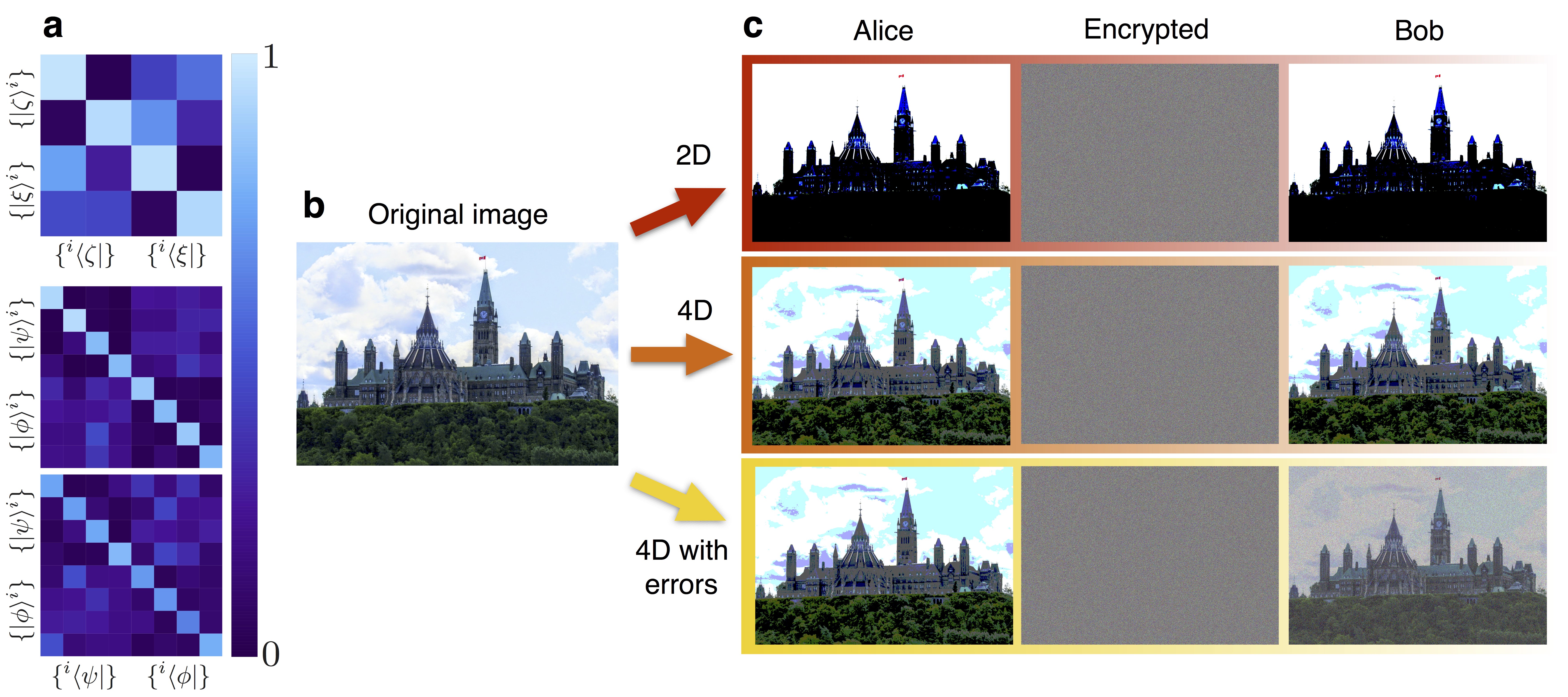}
	\caption[]{{\bf Experimental encryption of an image with structured photons.} {\bf a}, Probability-of-detection matrices, ${\cal P}^{i,j}=|{}^i\!\braket{\alpha}{\beta}^j|^2$ where $\alpha,\beta=\{\psi,\phi\}$, for 2D (top row), and 4D structured photons under different turbulence conditions (middle row: medium turbulence; bottom row: stronger turbulence). These matrices have the corresponding bit error rates of $Q^{2\mathrm{D}}=5\%$, $Q^{4\mathrm{D}}=11\%$, and $Q_{\mathrm{noisy}}^{4\mathrm{D}}=27\%$, respectively. {\bf b}, Image of the Parliament of Canada that Alice encrypts and sends to Bob through a classical channel using their shared secret key. {\bf c}, Using the experimentally measured probability of detection matrices ({\bf a}), Alice discretizes her intended image (left column) with $d$ levels, where $d$ is the encryption dimension, such that each pixel corresponds to three single photons (RGB values, leading to $d^3$ colours per pixel) that she sends to Bob. Alice then adds the shared secret key, generated from a BB84 protocol, on top of her discretized image to encrypt it (middle column). Bob decrypts Alice's sent image with his shared key to recover the image (left column). Implementing a 4-dimensional state clearly allows the ability to send more information per photon. The bit error rate is higher than the threshold of $Q_0^{4D}=18.9\%$ in the bottom row. Therefore, Bob cannot perform both privacy amplification and error correction when he decrypts the image. However, it is still low enough for him to perform privacy amplification, but the decrypted image is noisy as compared to the recovered image in the middle row where the error bit rate is below the threshold.}
	\label{fig:fig3}
	\end{center}
\end{figure*}
In the heralded single-photon source, the signal ($\lambda_s$ = 850~nm) and idler ($\lambda_i$ = 775~nm) photons are each coupled into separate single mode fibres (SMFs) to spatially filter them into their fundamental spatial mode. Alice takes the signal photon and prepares it in one of the states of the different MUBs through the use of an appropriate sequence of waveplates and $q$-plate. She then recombines the signal and idler photons on a dichroic mirror such that they are sent in the same beam across the link to Bob. It should be noted that despite sending two photons across the link simultaneously, our scheme is immune to photon-number-splitting attacks because the heralding photon \emph{does not} contain any of the polarization or OAM information of the signal photon. At the other end of the link, Bob receives the photon pairs and splits them at another dichroic mirror such that the idler photon is directly coupled into a SMF to act as a herald for the signal photon. With a sequence of waveplates, $q$-plate, PBS and SMF, mirror to that of Alice's, Bob can make a measurement on the signal photon by projecting it onto one of the states from one of the MUBs. In such a way, Bob has a spatial mode filter such that if he projects onto the same state that Alice sent, the signal photon will be phase-flattened and optimally detected. With avalanche photodiodes (APDs) connected to a coincidence logic box, the idler photon acts as a trigger for the arrival of the signal photon within a coincidence window of 5~ns, and the coincidence rates are recorded. The best performance of our free-space link after coupling to the SMFs on Bob's side gave count rates for the signal and idler photons of 0.75~MHz and 2.5~MHz, respectively, with an optimal coincidence rate of almost 50~kHz. However, due to large temperature and turbulence differences from night to night, the numbers varied throughout the various experimental runs. Overall, from sender to receiver, there are approximately 20\% and 25\% coupling efficiencies (equivalently 7~dB and 6~dB of losses) for the signal and idler photons, respectively, which gives an approximately 5\% success rate for recording coincidences. Since no adaptive optics was utilized, many of the raw data points sampled are badly perturbed by the turbulence. The most dominant effect of the atmospheric turbulence given the range of our atmospheric structure constant, $C_n^2$, between $2.5\times10^{-15}$~m$^{-2/3}$ and $6.4\times10^{-16}$~m$^{-2/3}$ (see Methods for details) is beam wandering~\cite{ageorges2013laser}. The idler photon is still in the fundamental mode and would always couple to the SMF under stable conditions, but does not optimally couple due to the turbulence. We thus use the idler photon as not only a herald for the signal photon but also as a ``target'' to gauge the beam wandering in Bob's setup which helps to correct our measurements for turbulence. The signal photon is co-axially propagating with the idler photon, and thus experiences the same atmospheric turbulence. Though not crucial, this method is effective and aids in decreasing the bit error rate.

In QKD, a secret key may be established between Alice and Bob with a secret key rate, defined as the number of bits of secret key established divided by the number of sifted photons, given by $R(Q)=\log_2(d)-2h(Q)$, where $Q$ is the quantum bit error rate and $h(\cdot)$ is the Shannon entropy in dimension $d$. Hence, there is a threshold value of $Q_0$ above which a non-zero shared secure key cannot be generated. In dimension 2, this threshold value is the well-known $Q_0^{2\mathrm{D}}=11.0\%$, while it almost doubles to $Q_0^{4\mathrm{D}}=18.9\%$, in dimension 4~\cite{cerf2002security}. This clearly exhibits the robustness of high-dimensional quantum cryptography. We perform a 4-dimensional BB84 protocol under different atmospheric conditions. Probability-of-detection matrices for the 4-dimensional structured photonic states, $\{ \ket{\psi}^i \}$ and $\{ \ket{\phi}^j \}$ with $\ell=2$, of the BB84 protocol are shown in Fig.~\ref{fig:fig3}a (middle row). In dimension 4, from the \emph{raw} probability-of-detection matrix, the quantum bit error rate is $Q=14\%$, and is below the threshold value of $Q_0^{4\mathrm{D}}$, resulting in a positive corresponding secret key rate of $R=0.39$~bits per sifted photon. By considering the idler as a target beam, which accounts for turbulence, the quantum bit error rate is reduced to $Q^{4\mathrm{D}}=11\%$ with a secret key rate of $R^{4\mathrm{D}}=0.65$~bits per sifted photon. The secret key rate is lower than the maximum theoretical value of 2~bits per sifted photon, which is due to imperfections in transmission. For a comparison, we perform a BB84 with two-dimensional structured photons in the MUBs of $\ket{\zeta}=\{ \left(\ket{L,-1}\pm \ket{R,1}\right)/\sqrt{2} \}$ and $\ket{\xi} = \{ \left( \ket{L,-1}\pm i\ket{R,1} \right) /\sqrt{2} \}$, see Fig.~\ref{fig:fig3}a (top row). A quantum bit error rate and secret key rate of $Q^{2\mathrm{D}}=5\%$ and $R^{2\mathrm{D}}=0.43$~bits per sifted photon were obtained, respectively, using the target as compensation. Indeed, $R^{4\mathrm{D}}$ is larger than $R^{2\mathrm{D}}$ showing the potential for transmitting more secure information per sifted photon in higher dimensions. This is visually shown in Fig.~\ref{fig:fig3}c (top and middle row): the image that Alice sends Bob (Fig.~\ref{fig:fig3}b) can be discretized with more steps in dimension 4 (middle row) as compared to dimension 2 (top row). Due to turbulence, the quantum bit error rate for dimension 4 on many nights was above $Q_0^{4\mathrm{D}}$. An example of one of these nights is shown in Fig.~\ref{fig:fig3}a (bottom row) with a calculated quantum bit error rate of $Q^{4\mathrm{D}}_{\mathrm{noisy}}=27\%$ calculated from the probability-of-detection matrix. Although secure, any image sent to Bob would appear noisy after decryption, shown in Fig.~\ref{fig:fig3}c (bottom row). However, allowing for two-way classical communications, the tolerable error bit rate increases to $31.5\% > Q^{4\mathrm{D}}_{\mathrm{noisy}}$ in dimension 4~\cite{nikolopoulos2006error} (see Supplementary Information). 

We have shown the feasibility of increasing the secure data transmission rate using high-dimensional quantum states compared to bi-dimensional states despite a noisy channel. This paves the road towards high-dimensional intra-city quantum cryptography via quantum key distribution. In addition, our results lay the groundwork for intra-city quantum teleportation with structured photons, which is an essential component of a free-space quantum network. These demonstrations can be extended over longer distances provided there is adequate turbulence monitoring and compensation.



\vspace{0.5cm}
\noindent\textbf{Methods}\newline

\noindent{\footnotesize{\bf Experimental Setup:} Single photon pairs are generated via the spontaneous parametric down-conversion (SPDC) process in a 5-mm-long ppKTP crystal pumped by a 405~nm laser diode (200~mW). Nondegenerate wavelengths for the signal ($\lambda_s$= 850~nm) and idler ($\lambda_i$=775~nm) photons are chosen in order to efficiently separate the two; only the signal photon is encoded with information. The signal and idler are each coupled into a separate single mode fibre (SMF) to spatially filter the photons into the fundamental mode. Bandpass filters, 850$\pm$5~nm and 775$\pm$20~nm, are placed in front of the fibre couplers to select the correct photon pairs. The singles count rates after the SMFs, detected with avalanche photodiodes (APDs), are 4~MHz and 10~MHz for the signal and idler, respectively. The idler photon heralds the presence of the signal photon, as determined by a coincidence logic box. This procedure gives a coincidence rate of around 1~MHz for a coincidence window of 5~ns with $\lesssim$0.2~MHz of accidental coincidence detections. After the encoding of the quantum information by wave plates and a $q$-plate on the signal photon, we recombine it with its partner photon by means of a dichroic mirror, and enlarge both spatial structures to minimize divergence upon propagation. At the last lens ($f_2$) of the sending unit the beam waist is approximately 12~mm. After propagation over the 0.3~km distance, we find the beam waist to be enlarged to approximately 20~mm as a consequence of atmospheric influences and imperfect optics. In order to measure the received quantum states, we demagnify the photon's structure with another set of lenses, separate the information-carrying signal photon from the heralding trigger photon with another dichroic mirror and measure its state, again with the help of waveplates and a $q$-plate. For more details about focal lengths, see main text and figure caption \ref{fig:fig2}.}\newline

{\noindent{\footnotesize{\bf Turbulence Characterization:} To characterize the Ottawa intra-city free space link, we investigate the turbulence by evaluating its characteristic properties such as the atmospheric structure constant $C_n^2$ and the Fried parameter $r_0$~\cite{kolmogorov1941local,fried1966optical,ageorges2013laser}. We do so by sending a Gaussian-shaped laser beam (850~nm) over the 0.3~km-long link and record its arrival position with a CCD camera. Because atmospheric turbulence changes on a millisecond time scale, short-term exposure images can reveal beam wandering, which is caused by fast-moving air cells, each having slightly different pressures, and thus small differences in refractive indices. The stronger the turbulence and the larger the distance of the link, the larger are the deflections from the optical axis. The latter can be deduced by taking an average over many short term exposure images, which effectively leads to an atmospherically broadened Gaussian beam profile. During different measurement nights, we record 500 short exposure images (0.07~ms each), from which we calculate a Fried parameter between 18~cm and 41~cm, which corresponds to an atmospheric structure constant $C_n^2$ ranging from around $2.5\times10^{-15}$~m$^{-2/3}$ to $6.4\times10^{-16}$~m$^{-2/3}$, assuming Kolmogorov theory for atmospheric turbulence. Hence, the link shows moderate turbulence effects on the transmitted light fields.}}


\begin{thebibliography}{10}
\expandafter\ifx\csname url\endcsname\relax
  \def\url#1{\texttt{#1}}\fi
\expandafter\ifx\csname urlprefix\endcsname\relax\def\urlprefix{URL }\fi
\providecommand{\bibinfo}[2]{#2}
\providecommand{\eprint}[2][]{\url{#2}}

\bibitem{scarani2009security}
\bibinfo{author}{Scarani, V.} \emph{et~al.}
\newblock \bibinfo{title}{The security of practical quantum key distribution}.
\newblock \emph{\bibinfo{journal}{Reviews of Modern Physics}}
  \textbf{\bibinfo{volume}{81}}, \bibinfo{pages}{1301} (\bibinfo{year}{2009}).

\bibitem{lo2014secure}
\bibinfo{author}{Lo, H.-K.}, \bibinfo{author}{Curty, M.} \&
  \bibinfo{author}{Tamaki, K.}
\newblock \bibinfo{title}{Secure quantum key distribution}.
\newblock \emph{\bibinfo{journal}{Nature Photonics}}
  \textbf{\bibinfo{volume}{8}}, \bibinfo{pages}{595--604}
  (\bibinfo{year}{2014}).

\bibitem{bechmann2000quantum}
\bibinfo{author}{Bechmann-Pasquinucci, H.} \& \bibinfo{author}{Tittel, W.}
\newblock \bibinfo{title}{Quantum cryptography using larger alphabets}.
\newblock \emph{\bibinfo{journal}{Physical Review A}}
  \textbf{\bibinfo{volume}{61}}, \bibinfo{pages}{062308}
  (\bibinfo{year}{2000}).

\bibitem{cerf2002security}
\bibinfo{author}{Cerf, N.~J.}, \bibinfo{author}{Bourennane, M.},
  \bibinfo{author}{Karlsson, A.} \& \bibinfo{author}{Gisin, N.}
\newblock \bibinfo{title}{Security of quantum key distribution using d-level
  systems}.
\newblock \emph{\bibinfo{journal}{Physical Review Letters}}
  \textbf{\bibinfo{volume}{88}}, \bibinfo{pages}{127902}
  (\bibinfo{year}{2002}).

\bibitem{groblacher2006experimental}
\bibinfo{author}{Gr{\"o}blacher, S.}, \bibinfo{author}{Jennewein, T.},
  \bibinfo{author}{Vaziri, A.}, \bibinfo{author}{Weihs, G.} \&
  \bibinfo{author}{Zeilinger, A.}
\newblock \bibinfo{title}{Experimental quantum cryptography with qutrits}.
\newblock \emph{\bibinfo{journal}{New Journal of Physics}}
  \textbf{\bibinfo{volume}{8}}, \bibinfo{pages}{75} (\bibinfo{year}{2006}).

\bibitem{mafu2013higher}
\bibinfo{author}{Mafu, M.} \emph{et~al.}
\newblock \bibinfo{title}{Higher-dimensional orbital-angular-momentum-based
  quantum key distribution with mutually unbiased bases}.
\newblock \emph{\bibinfo{journal}{Physical Review A}}
  \textbf{\bibinfo{volume}{88}}, \bibinfo{pages}{032305}
  (\bibinfo{year}{2013}).

\bibitem{mirhosseini2015high}
\bibinfo{author}{Mirhosseini, M.} \emph{et~al.}
\newblock \bibinfo{title}{High-dimensional quantum cryptography with twisted
  light}.
\newblock \emph{\bibinfo{journal}{New Journal of Physics}}
  \textbf{\bibinfo{volume}{17}}, \bibinfo{pages}{033033}
  (\bibinfo{year}{2015}).

\bibitem{nagali2010experimental}
\bibinfo{author}{Nagali, E.}, \bibinfo{author}{Sansoni, L.},
  \bibinfo{author}{Marrucci, L.}, \bibinfo{author}{Santamato, E.} \&
  \bibinfo{author}{Sciarrino, F.}
\newblock \bibinfo{title}{Experimental generation and characterization of
  single-photon hybrid ququarts based on polarization and orbital angular
  momentum encoding}.
\newblock \emph{\bibinfo{journal}{Physical Review A}}
  \textbf{\bibinfo{volume}{81}}, \bibinfo{pages}{052317}
  (\bibinfo{year}{2010}).

\bibitem{allen1992orbital}
\bibinfo{author}{Allen, L.}, \bibinfo{author}{Beijersbergen, M.~W.},
  \bibinfo{author}{Spreeuw, R.} \& \bibinfo{author}{Woerdman, J.}
\newblock \bibinfo{title}{Orbital angular momentum of light and the
  transformation of laguerre-gaussian laser modes}.
\newblock \emph{\bibinfo{journal}{Physical Review A}}
  \textbf{\bibinfo{volume}{45}}, \bibinfo{pages}{8185} (\bibinfo{year}{1992}).

\bibitem{mair2001entanglement}
\bibinfo{author}{Mair, A.}, \bibinfo{author}{Vaziri, A.},
  \bibinfo{author}{Weihs, G.} \& \bibinfo{author}{Zeilinger, A.}
\newblock \bibinfo{title}{Entanglement of the orbital angular momentum states
  of photons}.
\newblock \emph{\bibinfo{journal}{Nature}} \textbf{\bibinfo{volume}{412}},
  \bibinfo{pages}{313--316} (\bibinfo{year}{2001}).

\bibitem{molina:2007}
\bibinfo{author}{Molina-Terriza, G.}, \bibinfo{author}{Torres, J.~P.} \&
  \bibinfo{author}{Torner, L.}
\newblock \bibinfo{title}{Twisted photons}.
\newblock \emph{\bibinfo{journal}{Nature Physics}}
  \textbf{\bibinfo{volume}{3}}, \bibinfo{pages}{305--310}
  (\bibinfo{year}{2007}).

\bibitem{cardano2015quantum}
\bibinfo{author}{Cardano, F.} \emph{et~al.}
\newblock \bibinfo{title}{Quantum walks and wavepacket dynamics on a lattice
  with twisted photons}.
\newblock \emph{\bibinfo{journal}{Science Advances}}
  \textbf{\bibinfo{volume}{1}}, \bibinfo{pages}{e1500087}
  (\bibinfo{year}{2015}).

\bibitem{willner2015optical}
\bibinfo{author}{Willner, A.~E.} \emph{et~al.}
\newblock \bibinfo{title}{Optical communications using orbital angular momentum
  beams}.
\newblock \emph{\bibinfo{journal}{Advances in Optics and Photonics}}
  \textbf{\bibinfo{volume}{7}}, \bibinfo{pages}{66--106}
  (\bibinfo{year}{2015}).

\bibitem{paterson2005atmospheric}
\bibinfo{author}{Paterson, C.}
\newblock \bibinfo{title}{Atmospheric turbulence and orbital angular momentum
  of single photons for optical communication}.
\newblock \emph{\bibinfo{journal}{Physical review letters}}
  \textbf{\bibinfo{volume}{94}}, \bibinfo{pages}{153901}
  (\bibinfo{year}{2005}).

\bibitem{malik2012influence}
\bibinfo{author}{Malik, M.} \emph{et~al.}
\newblock \bibinfo{title}{Influence of atmospheric turbulence on optical
  communications using orbital angular momentum for encoding}.
\newblock \emph{\bibinfo{journal}{Optics Express}}
  \textbf{\bibinfo{volume}{20}}, \bibinfo{pages}{13195--13200}
  (\bibinfo{year}{2012}).

\bibitem{farias2015resilience}
\bibinfo{author}{Far{\'\i}as, O.~J.} \emph{et~al.}
\newblock \bibinfo{title}{Resilience of hybrid optical angular momentum qubits
  to turbulence}.
\newblock \emph{\bibinfo{journal}{Scientific reports}}
  \textbf{\bibinfo{volume}{5}} (\bibinfo{year}{2015}).

\bibitem{goyal2016effect}
\bibinfo{author}{Goyal, S.~K.}, \bibinfo{author}{Roux, F.~S.},
  \bibinfo{author}{Konrad, T.}, \bibinfo{author}{Forbes, A.} \emph{et~al.}
\newblock \bibinfo{title}{The effect of turbulence on entanglement-based
  free-space quantum key distribution with photonic orbital angular momentum}.
\newblock \emph{\bibinfo{journal}{Journal of Optics}}
  \textbf{\bibinfo{volume}{18}}, \bibinfo{pages}{064002}
  (\bibinfo{year}{2016}).

\bibitem{wang2012terabit}
\bibinfo{author}{Wang, J.} \emph{et~al.}
\newblock \bibinfo{title}{Terabit free-space data transmission employing
  orbital angular momentum multiplexing}.
\newblock \emph{\bibinfo{journal}{Nature Photonics}}
  \textbf{\bibinfo{volume}{6}}, \bibinfo{pages}{488--496}
  (\bibinfo{year}{2012}).

\bibitem{tamburini:2012}
\bibinfo{author}{Tamburini, F.} \emph{et~al.}
\newblock \bibinfo{title}{Encoding many channels on the same frequency through
  radio vorticity: first experimental test}.
\newblock \emph{\bibinfo{journal}{New Journal of Physics}}
  \textbf{\bibinfo{volume}{14}}, \bibinfo{pages}{033001}
  (\bibinfo{year}{2012}).

\bibitem{lavery:2015}
\bibinfo{author}{Lavery, M.~P.} \emph{et~al.}
\newblock \bibinfo{title}{Study of turbulence induced orbital angular momentum
  channel crosstalk in a 1.6 km free-space optical link}.
\newblock In \emph{\bibinfo{booktitle}{CLEO: Science and Innovations}},
  \bibinfo{pages}{STu1L--4} (\bibinfo{organization}{Optical Society of
  America}, \bibinfo{year}{2015}).

\bibitem{krenn:2014}
\bibinfo{author}{Krenn, M.} \emph{et~al.}
\newblock \bibinfo{title}{Communication with spatially modulated light through
  turbulent air across vienna}.
\newblock \emph{\bibinfo{journal}{New Journal of Physics}}
  \textbf{\bibinfo{volume}{16}}, \bibinfo{pages}{113028}
  (\bibinfo{year}{2014}).

\bibitem{Krenn2016Twisted}
\bibinfo{author}{Krenn, M.} \emph{et~al.}
\newblock \bibinfo{title}{Twisted light transmission over 143 km}.
\newblock \emph{\bibinfo{journal}{Proceedings of the National Academy of
  Sciences}} \textbf{\bibinfo{volume}{113}}, \bibinfo{pages}{13648--13653}
  (\bibinfo{year}{2016}).

\bibitem{vallone:2014}
\bibinfo{author}{Vallone, G.} \emph{et~al.}
\newblock \bibinfo{title}{Free-space quantum key distribution by
  rotation-invariant twisted photons}.
\newblock \emph{\bibinfo{journal}{Physical Review Letters}}
  \textbf{\bibinfo{volume}{113}}, \bibinfo{pages}{060503}
  (\bibinfo{year}{2014}).

\bibitem{krenn2015twisted}
\bibinfo{author}{Krenn, M.}, \bibinfo{author}{Handsteiner, J.},
  \bibinfo{author}{Fink, M.}, \bibinfo{author}{Fickler, R.} \&
  \bibinfo{author}{Zeilinger, A.}
\newblock \bibinfo{title}{Twisted photon entanglement through turbulent air
  across vienna}.
\newblock \emph{\bibinfo{journal}{Proceedings of the National Academy of
  Sciences}} \textbf{\bibinfo{volume}{112}}, \bibinfo{pages}{14197--14201}
  (\bibinfo{year}{2015}).

\bibitem{zhan2009cylindrical}
\bibinfo{author}{Zhan, Q.}
\newblock \bibinfo{title}{Cylindrical vector beams: from mathematical concepts
  to applications}.
\newblock \emph{\bibinfo{journal}{Advances in Optics and Photonics}}
  \textbf{\bibinfo{volume}{1}}, \bibinfo{pages}{1--57} (\bibinfo{year}{2009}).

\bibitem{bennett1984quantum}
\bibinfo{author}{Bennett, C.~H.} \& \bibinfo{author}{Brassard, G.}
\newblock \bibinfo{title}{Quantum cryptography: Public key distribution and
  coin tossing}.
\newblock In \emph{\bibinfo{booktitle}{International Conference on Computer
  System and Signal Processing, IEEE, 1984}}, \bibinfo{pages}{175--179}
  (\bibinfo{year}{1984}).

\bibitem{marrucci2006optical}
\bibinfo{author}{Marrucci, L.}, \bibinfo{author}{Manzo, C.} \&
  \bibinfo{author}{Paparo, D.}
\newblock \bibinfo{title}{Optical spin-to-orbital angular momentum conversion
  in inhomogeneous anisotropic media}.
\newblock \emph{\bibinfo{journal}{Physical Review Letters}}
  \textbf{\bibinfo{volume}{96}}, \bibinfo{pages}{163905}
  (\bibinfo{year}{2006}).

\bibitem{larocque2016arbitrary}
\bibinfo{author}{Larocque, H.} \emph{et~al.}
\newblock \bibinfo{title}{Arbitrary optical wavefront shaping via spin-to-orbit
  coupling}.
\newblock \emph{\bibinfo{journal}{Journal of Optics}}
  \textbf{\bibinfo{volume}{18}}, \bibinfo{pages}{124002}
  (\bibinfo{year}{2016}).

\bibitem{nagali2009quantum}
\bibinfo{author}{Nagali, E.} \emph{et~al.}
\newblock \bibinfo{title}{Quantum information transfer from spin to orbital
  angular momentum of photons}.
\newblock \emph{\bibinfo{journal}{Physical Review Letters}}
  \textbf{\bibinfo{volume}{103}}, \bibinfo{pages}{013601}
  (\bibinfo{year}{2009}).

\bibitem{ageorges2013laser}
\bibinfo{author}{Ageorges, N.} \& \bibinfo{author}{Dainty, C.}
\newblock \emph{\bibinfo{title}{Laser Guide Star Adaptive Optics for
  Astronomy}}, vol. \bibinfo{volume}{551} (\bibinfo{publisher}{Springer Science
  \& Business Media}, \bibinfo{year}{2013}).

\bibitem{nikolopoulos2006error}
\bibinfo{author}{Nikolopoulos, G.~M.}, \bibinfo{author}{Ranade, K.~S.} \&
  \bibinfo{author}{Alber, G.}
\newblock \bibinfo{title}{Error tolerance of two-basis quantum-key-distribution
  protocols using qudits and two-way classical communication}.
\newblock \emph{\bibinfo{journal}{Physical Review A}}
  \textbf{\bibinfo{volume}{73}}, \bibinfo{pages}{032325}
  (\bibinfo{year}{2006}).

\bibitem{kolmogorov1941local}
\bibinfo{author}{Kolmogorov, A.~N.}
\newblock \bibinfo{title}{The local structure of turbulence in incompressible
  viscous fluid for very large reynolds numbers}.
\newblock In \emph{\bibinfo{booktitle}{Dokl. Akad. Nauk SSSR}},
  vol.~\bibinfo{volume}{30}, \bibinfo{pages}{301--305}
  (\bibinfo{organization}{JSTOR}, \bibinfo{year}{1941}).

\bibitem{fried1966optical}
\bibinfo{author}{Fried, D.~L.}
\newblock \bibinfo{title}{Optical resolution through a randomly inhomogeneous
  medium for very long and very short exposures}.
\newblock \emph{\bibinfo{journal}{Journal of the Optical Society of America}}
  \textbf{\bibinfo{volume}{56}}, \bibinfo{pages}{1372--1379}
  (\bibinfo{year}{1966}).

\end{thebibliography}

\vspace{0.2cm}
\noindent\textbf{Supplementary Information} Supplementary text, and experimental data shown in Fig.~\ref{fig:fig3}a.
\vspace{0.5 EM}

\noindent\textbf{Author Contributions}
\noindent A.S, F.B, R.F., J. G-B. and E.K. designed and built the Ottawa intra-city link. A.S., F.B. and R.F. performed the experiment. H.L. and J.G-B. fabricated the $q$-plates. A.S., F.B. and R.F. analyzed the data. A.S., F.B., D.E., C.P., K.G., C.M., G.L., R.W.B. and E.K. designed the first iteration in Erlangen, Germany. K.H. developed the theory. C.M, G.L, R.W.B. and E.K supervised and conceived the idea. A.S., F.B., R.F., E.K. wrote the manuscript with the help of the other co-authors.
\vspace{0.5 EM}

\noindent\textbf{Acknowledgments}
\noindent All authors would like to thank Peter Banzer and Thomas Bauer for insightful discussions, and acknowledge the support of the Max Planck -- University of Ottawa Centre for Extreme and Quantum Photonics. E.K. would like to thank Guy LeBlanc, Donald Hopkins, David Needham, and Sean Kirkwood for their help on establishing the free-space link over the city of Ottawa. A.S. thanks Harold and Th\'er\`ese Sit for their continuous support and encouragement. The authors thank Norman Bouchard for providing the photo of the parliament of Canada. This work was supported by the Canada Research Chairs (CRC), and Canada Foundation for Innovation (CFI) Programs. A.S. and H.L. acknowledge the support of the Natural Sciences and Engineering Research Council of Canada (NSERC). F.B. acknowledges the support of the Vanier Canada Graduate Scholarships Program of the NSERC. R.F. acknowledges the support of the Banting postdoctoral fellowship of the NSERC. R.W.B acknowledges the support of the Canada Excellence Research Chairs (CERC) Program.
\vspace{0.5 EM}

\noindent\textbf{Author Information}
\noindent The authors declare no competing financial interests. Correspondence and requests for materials should be addressed to E.K. (ekarimi@uottawa.ca).

\renewcommand{\theequation}{SI\arabic{equation}}

\setcounter{table}{0}
\setcounter{equation}{0}
\setcounter{figure}{0}
\makeatletter
\renewcommand{\bibnumfmt}[1]{[SI#1]}
\renewcommand{\citenumfont}[1]{SI#1}
\renewcommand{\figurename}[1]{FIG. SI#1}

\clearpage
\onecolumngrid
\section*{{\Large Supplementary Information for}\\ High-Dimensional Intra-City Quantum Cryptography with Structured Photons}

\section*{\underline{\large{Part 1:}} Mutually Unbiased Basis}

Given a set of bases $\alpha_0,\ldots,\alpha_n$ of dimension $d$, they are said to be mutually unbiased with respect to one another if they satisfy the following condition,
\begin{equation} \label{Eq:mat}
	|^j\!\braket{\alpha_i}{\alpha_{i'}}^{j'}|^2 = \begin{cases}
			\delta_{j,j'}   & \forall~ i = i' \\
			\tfrac{1}{d} & \forall~ i \neq i' \\
	              \end{cases}
	; \hspace{2mm} i\in\{0,1,...n\}, \hspace{1mm} j\in\{1,2,..,d\}.
\end{equation}
For dimensions where $d$ is a power of a prime, $d+1$ mutually unbiased bases (MUBs) can be found. For 2-dimensional quantum key distribution (QKD) protocols, photons can be encoded using polarization and orbital angular momentum (OAM). We represent states of light that have a particular polarization and OAM value using a compound ket notation. In this way, if a photon has a certain polarization $\Pi$ and carries $\ell$ units of OAM, it is written as $\ket{\Pi,\ell}$.
\begin{figure}[h]
	\begin{center} \includegraphics[width=0.5\columnwidth]{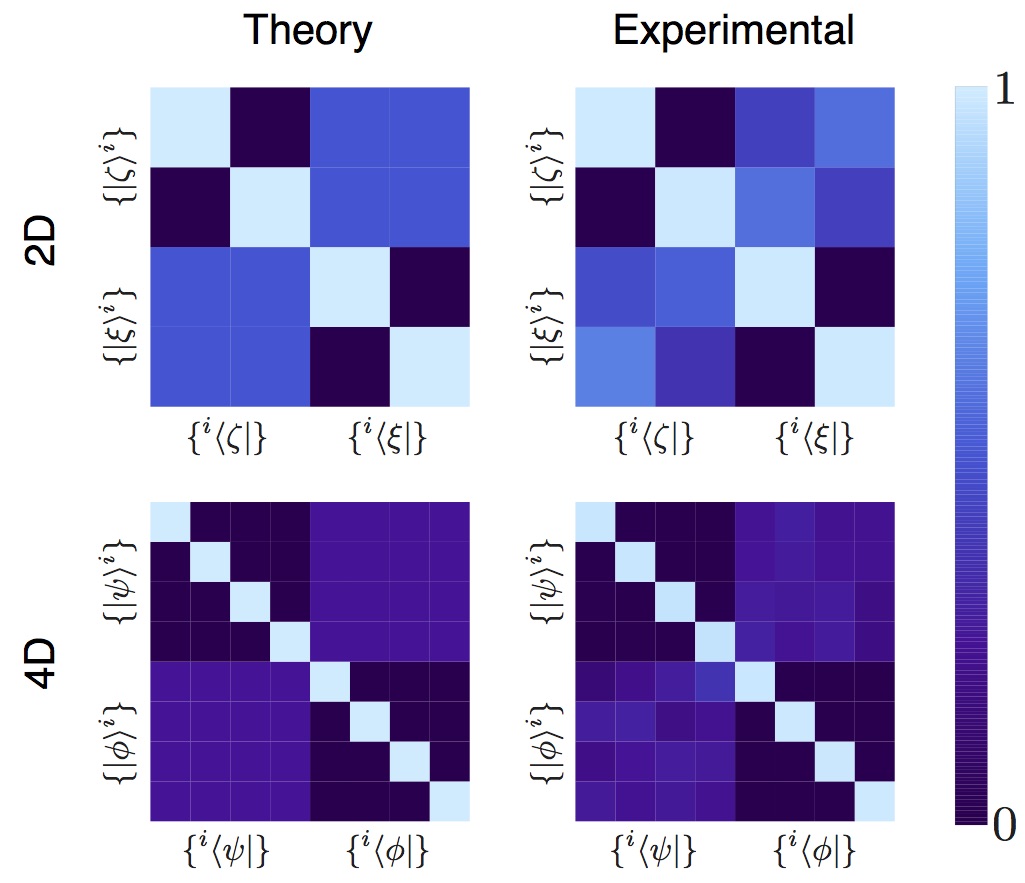}
	\caption{\textbf{Visualization of MUBs in d=2 and d=4} Theoretical probability-of-detection matrices (left column) for dimensions 2 and 4 using Eq.~(\ref{Eqs:MUBs}) and Eqs.~(\ref{eq:4DM0}--\ref{eq:4DM1}) by applying Eq.~(\ref{Eq:mat}). The probability-of-detection matrices as measured in the laboratory (right column) give bit error rates of 0.83\% and 1.83\% in dimensions 2 ($\ell=2$) and 4 ($\ell=2$), respectively.}
	\label{fig:SI}
	\end{center}
\end{figure}

The two MUBs of dimension 2 are given by,
\begin{eqnarray}\label{Eqs:MUBs}\nonumber
	\{ \ket{\zeta}^i \} &=& \left\{ \frac{1}{\sqrt{2}} \left(\ket{L,-\ell} + \ket{R,+\ell}\right), \frac{1}{\sqrt{2}} \left(\ket{L,-\ell} - \ket{R,+\ell}\right) \right\},
	\\ \{ \ket{\xi}^j \} &=& \left\{ \frac{1}{\sqrt{2}} \left(\ket{L,-\ell} + i\ket{R,+\ell}\right), \frac{1}{\sqrt{2}} \left(\ket{L,-\ell} - i\ket{R,+\ell}\right) \right\}.
\end{eqnarray}
In dimension 4, the natural basis is ${\ket{k}}\in\left\{ \ket{H,\ell}, \ket{H,-\ell}, \ket{V,\ell}, \ket{V,-\ell} \right\}$, and the two sets of MUBs $\{\ket{\psi}^i\}$ and $\{\ket{\phi}^j\}$ were generated by the following matrices,
\begin{eqnarray}\label{eq:sim} \nonumber
{\cal M}_0^{ik}&=&\left(\begin{array}{cccc}1 & 0 & 0 & 0 \\0 & 1 & 0 & 0 \\0 & 0 & 1 & 0 \\0 & 0 & 0 & 1\end{array}\right), \\
{\cal M}_1^{jk}&=&\frac{1}{2}\left(\begin{array}{cccc}1 & i & 1 & -i \\1 & i & -1 & i \\1 & -i & 1 & i \\-1 & i & 1 & i\end{array}\right).
\end{eqnarray}
such that $\ket{\psi}^i ={\cal M}_0^{ik}\ket{k}$ and $\ket{\phi}^j={\cal M}_1^{jk}\ket{k}$. This results in the following states:
\begin{eqnarray}
		\{ \ket{\psi}^i \} &=& \left\{ \ket{H,+\ell}, \ket{H,-\ell}, \ket{V,+\ell}, \ket{V,-\ell}  \right\}, \label{eq:4DM0}\\
		\{ \ket{\phi}^j \} &=& \left\{ \frac{1}{\sqrt{2}}(\ket{L,\ell}+\ket{R,-\ell}), \frac{1}{\sqrt{2}}(\ket{L,\ell}-\ket{R,-\ell}),\frac{1}{\sqrt{2}}(\ket{L,-\ell}+\ket{R,\ell}), \frac{1}{\sqrt{2}}(\ket{L,-\ell}-\ket{R,\ell})   \right\}. \label{eq:4DM1}
\end{eqnarray}
Figure SI1 shows a visual representation of the 2D (top row) and 4D (bottom row) MUBs using Eq.~(\ref{Eq:mat}), comparing the theoretical probability-of-detection matrix to the experimental one as measured in the laboratory, i.e. without the intra-city link. The quantum bit error rate is calculated as one minus the average of the on-diagonal elements. The calculated quantum bit error rates from the experimentally measured matrices are 0.83\% and 1.83\% in dimensions 2 ($\ell=2$) and 4 ($\ell=2$), respectively.

\section*{\underline{\large{Part 2:}} Generation of implemented MUBs in dimensions 2 and 4}
The MUBs in dimension 2, $\{ \ket{\zeta}^i \}$ and $\{ \ket{\xi}^j \}$ are generated using the sequence of a half-wave plate (HWP) followed by a $q$-plate, such that $\ell=2q$, where $q$ is the topological charge of the liquid crystal distribution. The waveplate angles are given in (\ref{table:table1}).
\begin{eqnarray} \label{table:table1}
	\begin{array}{c|c}
	\text{state} & \text{HWP} \\\hline 
	\ket{\zeta}^1 & 0^{\circ}\\\hline 
	\ket{\zeta}^2 & +45^{\circ} \\\hline 
	\ket{\xi}^1 & +22.5^{\circ} \\\hline 
	\ket{\xi}^2 & -22.5^{\circ}
	\end{array}
\end{eqnarray}

The MUBs in dimension 4, $\{ \ket{\psi}^i \}$ and $\{ \ket{\phi}^j \}$ are generated by sandwiching a $q$-plate between either HWPs or QWPs. If photons pass left to right through the following optical elements, the waveplate angles that Alice uses to generate $\{ \ket{\psi}^i \}$ are given in the (\ref{table:table2}), and $\{ \ket{\phi}^j \}$ in (\ref{table:table3}).
\begin{eqnarray} \label{table:table2}
	\begin{array}{c|c|c}
	\text{state} & \text{QWP before QP} &  \text{QWP after QP} \\\hline 
	\ket{\psi}^1 & -45^{\circ} & -45^{\circ} \\\hline 
	\ket{\psi}^2 & +45^{\circ} & +45^{\circ} \\\hline 
	\ket{\psi}^3 & -45^{\circ} & +45^{\circ} \\\hline 
	\ket{\psi}^4 & +45^{\circ} & -45^{\circ}
	\end{array}
\end{eqnarray}
\begin{eqnarray} \label{table:table3}
	\begin{array}{c|c|c}
	\text{state} & \text{HWP before QP} &  \text{HWP after QP} \\\hline 
	\ket{\phi}^1 & 0^{\circ} & 0^{\circ} \\\hline 
	\ket{\phi}^2 & +45^{\circ} & 0^{\circ} \\\hline 
	\ket{\phi}^3 & 0^{\circ} & - \\\hline 
	\ket{\phi}^4 & +45^{\circ} & -
	\end{array}
\end{eqnarray}
Bob uses the same waveplate angles, but mirrors the sequence of waveplates as Alice has in order to project his received photons onto a particular state.

\section*{\underline{\large{Part 3:}} Experimental data}
Coincidence counts are accumulated per 200~ms. For each of Bob's measurements, he records fifty data points. Bob obtains a probability-of-detection matrix by averaging the data points for each measurement and then normalizing over each state that Alice sends. The states that Alice sends are labelled and the states that Bob projects onto are labelled on the left and top, respectively, of each matrix below. 

Normalized raw data for probability of detection matrix in dimension 2 as measured across the intra-city link using a $q$=1/2 plate, as shown in Fig.~\ref{fig:fig3}a (top row):
\begin{eqnarray} \label{mat:2raw}
\bordermatrix{~	&	{}^1\!\bra{\zeta}	&	{}^2\!\bra{\zeta}	&\vr	{}^1\!\bra{\xi}	&	{}^2\!\bra{\xi}	\cr
\ket{\zeta}^1	&	0.971	& 	0.029	&\VR	0.421	&	0.579	\cr
\ket{\zeta}^2	&	0.062	&	0.938	&\VR	0.677	&	0.323	\cr\hline
\ket{\xi}^1	&	0.731	&	0.269	&\VR	0.959	&	0.041	\cr
\ket{\xi}^2	&	0.459	&	0.541	&\VR	0.068	&	0.932	\cr}
\end{eqnarray}

Target corrected data from (\ref{mat:2raw}):
\begin{eqnarray}
\bordermatrix{~	&	{}^1\!\bra{\zeta}	&	{}^2\!\bra{\zeta}	&\vr	{}^1\!\bra{\xi}	&	{}^2\!\bra{\xi}	\cr
\ket{\zeta}^1	&	0.972	&	0.028	&\VR	0.351	&	0.649	\cr
\ket{\zeta}^2	&	0.050	&	0.950	&\VR	0.653	&	0.347	\cr\hline
\ket{\xi}^1	&	0.725	&	0.275	&\VR	0.961	&	0.039	\cr
\ket{\xi}^2	&	0.463	&	0.537	&\VR	0.069	&	0.931	\cr
}
\end{eqnarray}
%

Normalized raw data for probability of detection matrix in dimension 4 as measured across the intra-city link:
\begin{eqnarray} \label{mat:4raw}
\bordermatrix{~	&	{}^1\!\bra{\psi}	&	{}^3\!\bra{\psi}	&	{}^2\!\bra{\psi}	&	{}^4\!\bra{\psi}	&\vr	{}^1\!\bra{\phi}	&	{}^2\!\bra{\phi}	&	{}^3\!\bra{\phi}^3	&	{}^4\!\bra{\phi}	\cr
\ket{\psi}^1	&	0.918	&	0.019	&	0.051	&	0.012	&\VR	0.252	&	0.245	&	0.275	&	0.228	\cr
\ket{\psi}^3	&	0.020	&	0.937	&	0.038	&	0.005	&\VR	0.190	&	0.192	&	0.312	&	0.306	\cr
\ket{\psi}^2	&	0.012	&	0.156	&	0.816	&	0.012	&\VR	0.279	&	0.277	&	0.289	&	0.155	\cr
\ket{\psi}^4	&	0.149	&	0.009	&	0.018	&	0.824	&\VR	0.152	&	0.195	&	0.384	&	0.269	\cr\hline
\ket{\phi}^1	&	0.319	&	0.125	&	0.325	&	0.231	&\VR	0.869	&	0.039	&	0.064	&	0.029	\cr
\ket{\phi}^2	&	0.252	&	0.217	&	0.239	&	0.292	&\VR	0.038	&	0.822	&	0.042	&	0.098	\cr
\ket{\phi}^3	&	0.185	&	0.177	&	0.447	&	0.191	&\VR	0.065	&	0.027	&	0.872	&	0.037	\cr
\ket{\phi}^4	&	0.207	&	0.205	&	0.381	&	0.208	&\VR	0.030	&	0.134	&	0.036	&	0.800	\cr
}
\end{eqnarray}

Target corrected data from (\ref{mat:4raw}), as shown in Fig.~\ref{fig:fig3}a (middle row):
\begin{eqnarray}
\bordermatrix{~	&	{}^1\!\bra{\psi}	&	{}^3\!\bra{\psi}	&	{}^2\!\bra{\psi}	&	{}^4\!\bra{\psi}	&\vr	{}^1\!\bra{\phi}	&	{}^2\!\bra{\phi}	&	{}^3\!\bra{\phi}^3	&	{}^4\!\bra{\phi} \cr
\ket{\psi}^1	&	0.924	&	0.035	&	0.011	&	0.031	&\VR	0.272	&	0.232	&	0.254	&	0.243	\cr
\ket{\psi}^3	&	0.024	&	0.960	&	0.012	&	0.004	&\VR	0.197	&	0.213	&	0.260	&	0.330	\cr
\ket{\psi}^2	&	0.005	&	0.052	&	0.930	&	0.013	&\VR	0.239	&	0.301	&	0.301	&	0.159	\cr
\ket{\psi}^4	&	0.049	&	0.004	&	0.029	&	0.918	&\VR	0.094	&	0.242	&	0.433	&	0.232	\cr\hline
\ket{\phi}^1	&	0.376	&	0.108	&	0.321	&	0.195	&\VR	0.874	&	0.033	&	0.065	&	0.028	\cr
\ket{\phi}^2	&	0.273	&	0.197	&	0.255	&	0.275	&\VR	0.035	&	0.825	&	0.045	&	0.096	\cr
\ket{\phi}^3	&	0.200	&	0.132	&	0.511	&	0.157	&\VR	0.060	&	0.016	&	0.889	&	0.035	\cr
\ket{\phi}^4	&	0.186	&	0.163	&	0.365	&	0.287	&\VR	0.026	&	0.129	&	0.043	&	0.803	\cr
}
\end{eqnarray}

Normalized raw data for probability of detection matrix in dimension 4 on a turbulent night as shown in Fig.~\ref{fig:fig3}a (bottom row):
\begin{eqnarray}
\bordermatrix{~	&	{}^1\!\bra{\psi}	&	{}^3\!\bra{\psi}	&	{}^2\!\bra{\psi}	&	{}^4\!\bra{\psi}	&\vr	{}^1\!\bra{\phi}	&	{}^2\!\bra{\phi}	&	{}^3\!\bra{\phi}^3	&	{}^4\!\bra{\phi} \cr
\ket{\psi}^1	&	0.741	&	0.032	&	0.043	&	0.184	&\VR	0.370	&	0.168	&	0.364	&	0.098	\cr
\ket{\psi}^3	&	0.096	&	0.722	&	0.138	&	0.044	&\VR	0.120	&	0.432	&	0.221	&	0.228	\cr
\ket{\psi}^2	&	0.043	&	0.177	&	0.755	&	0.025	&\VR	0.276	&	0.247	&	0.197	&	0.281	\cr
\ket{\psi}^4	&	0.101	&	0.041	&	0.047	&	0.811	&\VR	0.122	&	0.433	&	0.332	&	0.113	\cr\hline
\ket{\phi}^1	&	0.126	&	0.471	&	0.197	&	0.206	&\VR	0.707	&	0.051	&	0.144	&	0.098	\cr
\ket{\phi}^2	&	0.211	&	0.234	&	0.352	&	0.203	&\VR	0.110	&	0.694	&	0.079	&	0.117	\cr
\ket{\phi}^3	&	0.265	&	0.285	&	0.259	&	0.191	&\VR	0.195	&	0.056	&	0.632	&	0.117	\cr
\ket{\phi}^4	&	0.478	&	0.146	&	0.185	&	0.191	&\VR	0.048	&	0.103	&	0.075	&	0.775	\cr
}
\end{eqnarray}

\section*{\underline{\large{Part 4:}} Numerical approach for the secret key rate calculation}
Here we use a numerical approach to calculate the secret key rate for the MUBs in the current experiment that are shown in Eqs.~(\ref{eq:sim}--\ref{eq:4DM1}). The secret key rate calculation below relies on the dual optimization problem that has recently been introduced as an efficient numerical approach for unstructured quantum key distribution~[SI1]. The main result in~[SI1] indicates that the achievable secure key rate is lower bounded by the following maximization problem,

%
\begin{eqnarray}\label{dual}
	K \geq \frac{\Theta}{\text{ln}2} - H(Z_A|Z_B),
\end{eqnarray}
where
\begin{eqnarray}
	\Theta := \max_{\overrightarrow{\lambda}}\left(-\left|\left| \sum_{j}Z_A^jR(\overrightarrow{\lambda})Z_A^j\right|\right|  -\overrightarrow{\lambda} \cdot \overrightarrow{\gamma} \right),
\end{eqnarray}
and
\begin{eqnarray}
	R(\overrightarrow{\lambda}) := \exp\left(-\mathbb{1} - \overrightarrow{\lambda}\cdot\overrightarrow{\Gamma}\right).
\end{eqnarray}

Here $Z_A$ ($Z_B$) denotes the measurement performed by Alice (Bob) to derive the raw key, and $\overrightarrow{\gamma}=\{\gamma_i :=\text{Tr}(\rho_{AB}\Gamma_i)\}$ are determined through average value of experimental measurements. 

For the generalized BB84 in dimension $d=4$ with two MUBs, the experimental constraints can be summarized to 
\begin{equation}
	\text{Key-map POVM:}\quad Z_A = \left\{|\psi\rangle^i\langle \psi|, \text{for}\quad i=1\cdots d=4\right\}
\end{equation}
\begin{equation}
	\text{Constraints:}\quad \langle \mathbb{1}\rangle = 1
\end{equation}
\begin{equation}
	\langle \mathbf{E_X}\rangle = Q
\end{equation}
\begin{equation}
	\langle \mathbf{E_Z}\rangle = Q
\end{equation}
where $\mathbf{E_{Z\,(X)}}$ are coarse-grained error operators in $\mathcal{M}_{0\,(1)}$ MUBs and defined as
\begin{equation}
	\mathbf{E_X} =  \mathbb{1} - \sum_{i}^{d=4}|\psi\rangle^i\langle \psi|\otimes|\psi\rangle^i\langle \psi|
\end{equation}
\begin{equation}
	\mathbf{E_Z} =  \mathbb{1} - \sum_{i}^{d=4}|\varphi\rangle^i\langle \varphi|\otimes|\varphi\rangle^i\langle \varphi|.
\end{equation}
Eqs.~(\ref{eq:4DM0}) and (\ref{eq:4DM1}) show the definition for $|\psi\rangle^i$ and $|\varphi\rangle^i$ basis states.

Figure SI2 shows the numerical result of the optimization problem in Eq.~(\ref{dual}) with MUBs in Eqs.~(\ref{eq:4DM0},\ref{eq:4DM1}) in comparison with the theoretical key rates in~[SI2-SI3]. This numerical approach may be extended to find secret key rate per signal with two-way classical communications to tolerate higher qubit error rates [SI4].
\begin{figure}[h]
	\begin{center} \includegraphics[width=0.6\columnwidth]{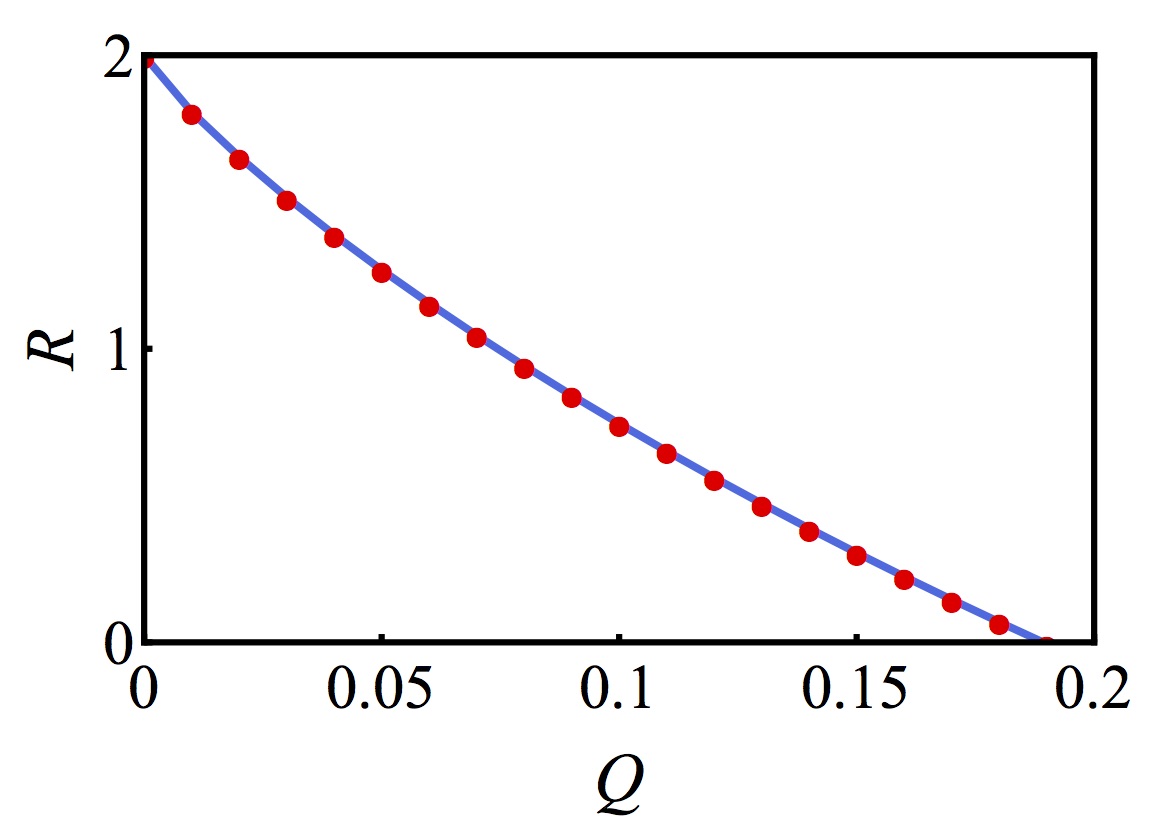}
	\caption{\textbf{Secret key rate per signal for BB84 in d=4 with 2 MUBs} Solution to the numerical optimization problem in Eq.~(\ref{dual}) are shown for different values of average error rates (red dots). As it can be seen, the numerical optimization saturates the bound and shows a good agreement with the theory from~[SI2,SI3]. For more details on the numerical approach see~[SI1].}
	\label{fig:SI}
	\end{center}
\end{figure}

\begin{itemize}
\item [][SI1] Patrick J. Coles, Eric M. Metodiev, and Norbert L\"utkenhaus, \textit{Nature Communications} {\bf 7}, 11712 (2016).
\item [][SI2] Agnes Ferenczi and Norbert L\"utkenhaus, \textit{Physical Review A} {\bf 85}, 052310 (2012).
\item [][SI3] L. Sheridan and V. Scarani, \textit{Physical Review A} {\bf 82}, 030301 (2010).
\item [][SI4] G. M. Nikolopoulos, K. S. Ranade and G. Alber, \textit{Physical Review A} {\bf 73}, 032325 (2006).
\end{itemize}

%
%
%
%
\end{document}